\documentclass[twocolumn, unsortedaddress, superscriptaddress]{revtex4}

\usepackage{siunitx}
\usepackage{tabularx}
\usepackage{graphicx}
\usepackage[colorlinks]{hyperref}
\usepackage{xcolor}

\usepackage{algorithm}
\usepackage{algpseudocode}

\definecolor{linkcolor}{RGB}{0,83,166}
\hypersetup{
  colorlinks = true,
  allcolors = {linkcolor}
}

\begin{document}
\newcommand{\mytitle}{Comparing Quantum Annealing and BF-DCQO}
\newcommand{\affildw}{D-Wave Quantum, Burnaby, British Columbia, Canada}

\author{Pau Farr\'{e}}
\email[]{pfarre@dwavesys.com}
\affiliation{\affildw}
\author{Erika Ordog}
\affiliation{\affildw}
\author{Kevin Chern}
\affiliation{\affildw}
\author{Catherine~C.~McGeoch}
\affiliation{D-Wave Quantum (Retired)}

\title{\mytitle}

\date{\today}
\begin{abstract}

Recent work \cite{Romero2025} has claimed that a gate-model 
quantum-classical hybrid algorithm called bias-field digitized counterdiabatic quantum optimization (BF-DCQO) \cite{Cadavid_2025} outperforms D-Wave's annealing quantum computers in optimization tasks. We find the opposite to be true, and demonstrate that D-Wave's quantum annealers find solutions of far greater quality than claimed in Ref.~\cite{Romero2025}, while using far less computation time. We also present evidence that suggests the quantum component of the hybrid approach makes minimal contributions to solution quality.

\end{abstract}

\maketitle

\def\title#1{\gdef\@title{#1}\gdef\THETITLE{#1}}

Quantum annealing (QA) is a computational paradigm where a quantum system stays near 
its instantaneous ground state while the Hamiltonian that governs its evolution is modified \cite{kadowaki,farhi2000}. The system is initialized in the
known ground state of a simple-to-prepare initial Hamiltonian, and then the Hamiltonian is slowly modified into a final Hamiltonian whose ground state
encodes the solution of an optimization problem.

D-Wave's quantum annealing processing units (QPU) implement the QA paradigm. The native problem implemented on these platforms contains linear and quadratic terms and is known as the Ising model or quadratic unconstrained binary optimization (QUBO).    

Gate model (GM) quantum computers can implement a discretized approximation to QA called quantum approximate optimization algorithm (QAOA) \cite{farhi2014}.  This approach uses a quantum circuit subroutine that is iteratively optimized, as a pair of parameter vectors ($\vec{\gamma}$, $\vec{\beta}$) is varied. The task of choosing the next parameters during the iteration is performed in a feedback loop that combines quantum and classical processing.  

Recently, Cadavid et al. \cite{Cadavid_2025} described the bias-field digitized counterdiabatic quantum optimization (BF-DCQO) method. This approach is similar to QAOA, with fixed ($\vec{\gamma}$, $\vec{\beta}$) and counterdiabatic terms \cite{Hegade_2022}. The approach also includes per-qubit bias fields (BF). These biases point toward the average value of the best solutions in the previous iteration. In some implementations, the classical component adds a greedy-descent refinement step to those solutions \cite{simen,chandarana,romero}. As a result, BF-DCQO poses a stark contrast to QAOA by letting the classical loop directly influence individual qubit values. This makes it possible for the classical solution refinement, selection, and biasing mechanisms to become the primary drivers for the hybrid solver's solution quality, possibly pushing the quantum computer to a secondary role in the hybrid approach in which it makes minimal contributions to solution quality.

Indeed, the authors of Ref. \cite{Cadavid_2025} report that when BF-DCQO experiments used a quantum computer, most iterations were performed on a classical simulator, and only their final iteration was performed on the quantum processors built by IonQ and IBM. The authors of Ref. \cite{Romero2025} similarly remark that an IBM processor was only applied to the final iteration of BF-DCQO.

\begin{figure}[t!] 
  \centering 
  \includegraphics[width=\columnwidth]{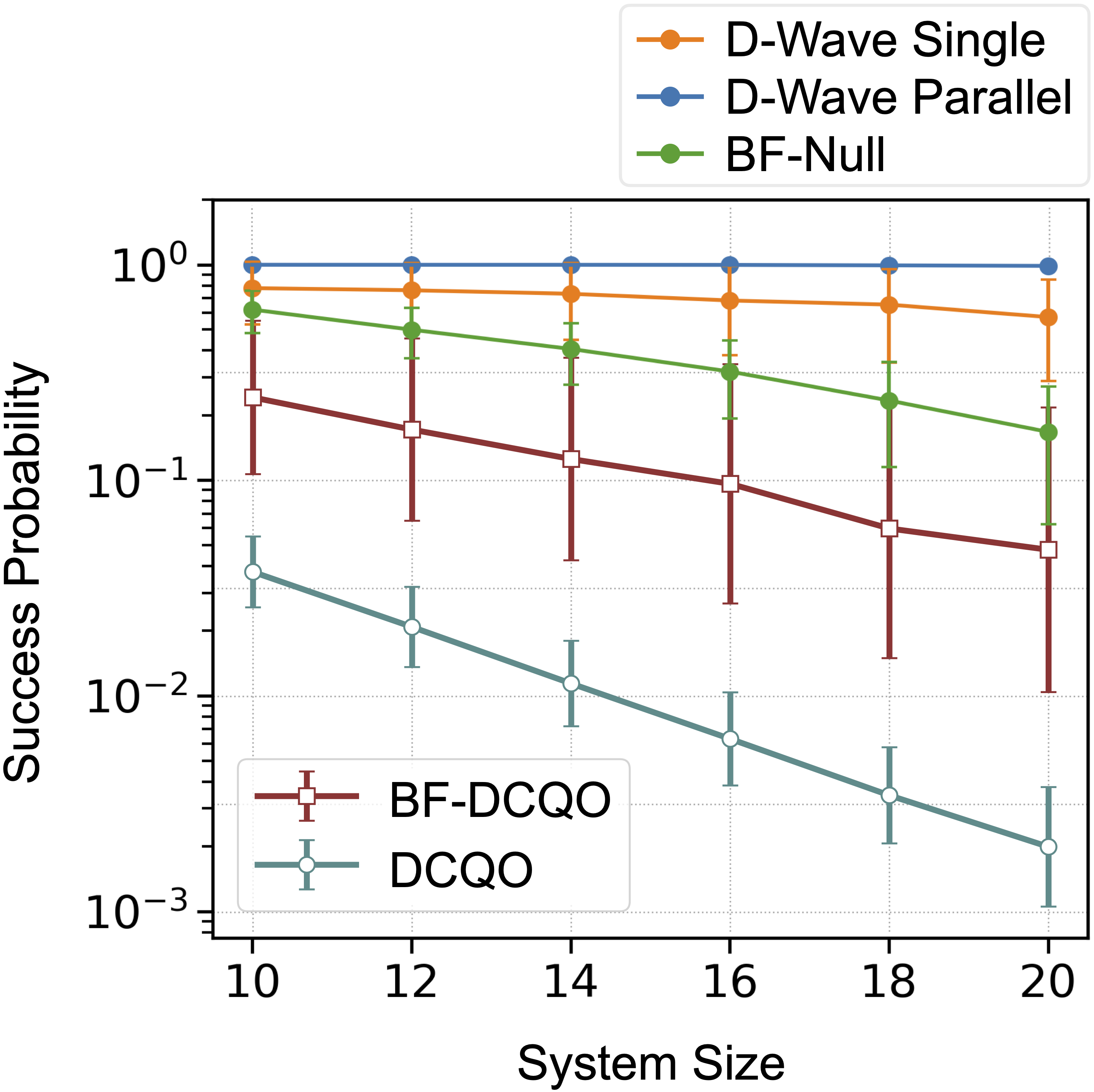} 
  \caption{Probability of ground state on ensembles of 400 random instances of the fully connected Ising model with Gaussian weights. BF-DCQO and DCQO results are reported in FIG.~1B in \cite{Cadavid_2025}.} 
  \label{fig:p_gs} 
\end{figure}

These reports prompt us to wonder whether the authors checked for the experimental null hypothesis: \textit{is the quantum component of BF-DCQO performing any non-trivial work?} 

We tested this hypothesis by comparing the performance of BF-DCQO with that of a dummy classical solver, the bias-field null-hypothesis algorithm (BF-Null) (Section \ref{subsec:bfnull}, Alg.\ref{alg:bfnh}). We show that substituting a poor classical solver for the quantum solver in the BF-DCQO workflow yields equal or better results than those reported in \cite{Cadavid_2025,Romero2025}. This result supports the null hypothesis that the contribution of the quantum computation to BF-DCQO is minimal.

\begin{figure}[t!] 
  \centering 
  \includegraphics[width=\columnwidth]{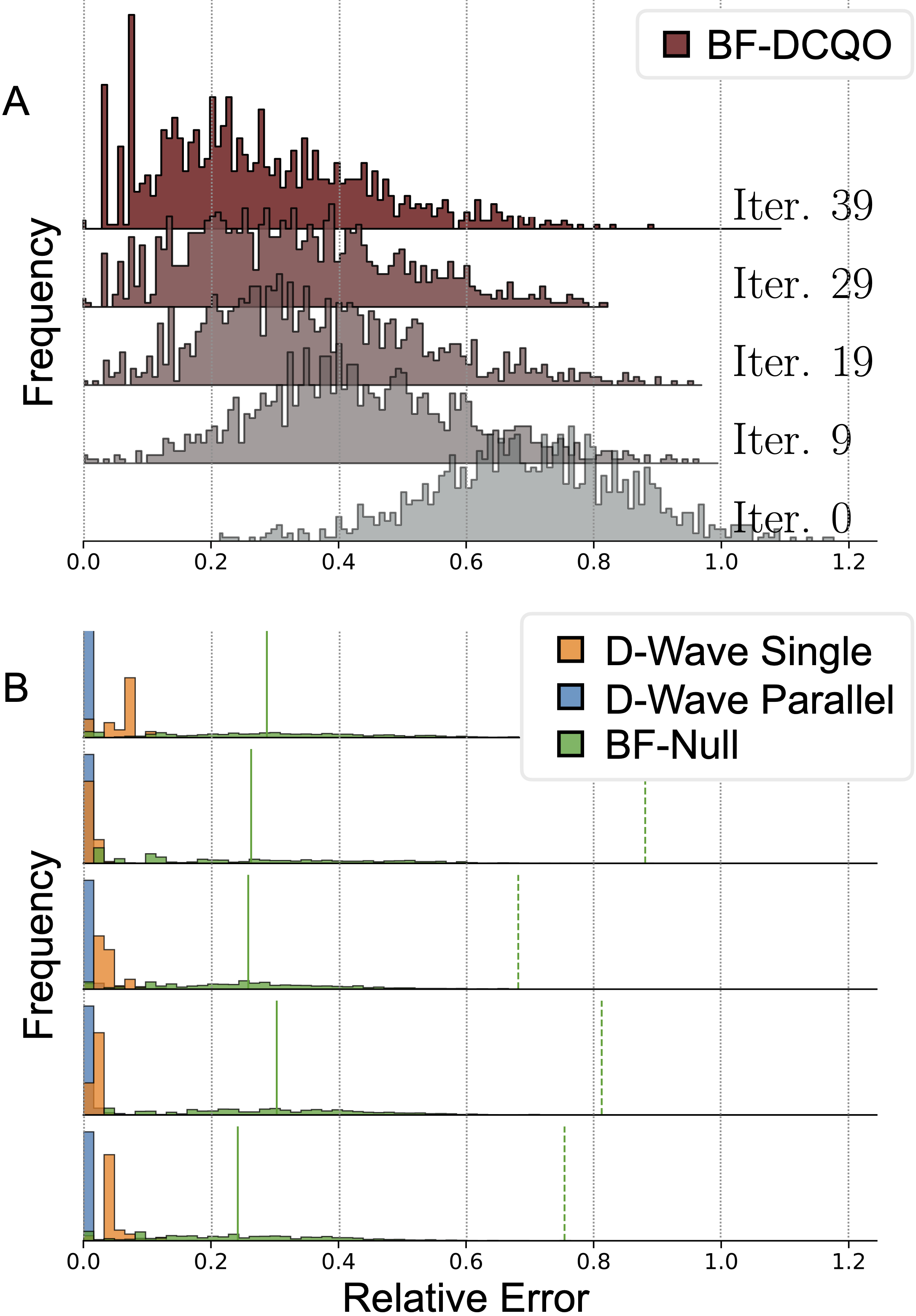} 
  \caption{Solution quality of 1000 reads for 29 qubit instances. (A) Distributions from BF-DCQO on the IonQ Forte noisy simulator over multiple iterations on a single instance, reported in FIG. 1D in \cite{Cadavid_2025}. (B) Solution qualities for five random instances run on the D-Wave quantum annealer using single and parallel embeddings and BF-Null (39 iterations). The solid and dashed green vertical lines highlight the mean and maximum of the NF-Null distribution for visibility.} 
  \label{fig:distA} 
\end{figure}

By the time of writing the first version of this paper, we had not received a response to our request for the data used in the two BF-DCQO studies that report comparisons to D-Wave\texttrademark~quantum processors \cite{Romero2025,simen}. We therefore used the problem definitions in \cite{Cadavid_2025,Romero2025} to build random generators for two input classes, called Ising and hising (i.e., higher-order Ising problem), to support comparison of the reported performance of BF-DQCO with results from our tests using D-Wave quantum processors. The lack of the details in \cite{simen} prevented us from generating the instances used in that study, so we did not attempt to replicate those results.

A year after our request, we obtained the specific hising instance used in \cite{Romero2025} to compare BF-DCQO to D-Wave. We report our updated results in Section~\ref{hising}. The ensemble of Ising instances in Section \ref{sec:Ising} continues to be generated randomly, but we note that these instances are statistically consistent with the reported ground states in \cite{Cadavid_2025}, and we believe they are accurate enough to support the conclusions of this paper.

Section \ref{sec:Results} presents experimental results showing that quantum annealing significantly outperforms BF-DQCO, returning far better-quality solutions than reported in Ref.~\cite{Cadavid_2025}. Section \ref{sec:timing} shows that D-Wave QPU runtimes are far shorter than estimated runtimes for BF-DCQO.

\begin{figure}[t!] 
  \centering 
  \includegraphics[width=\columnwidth]{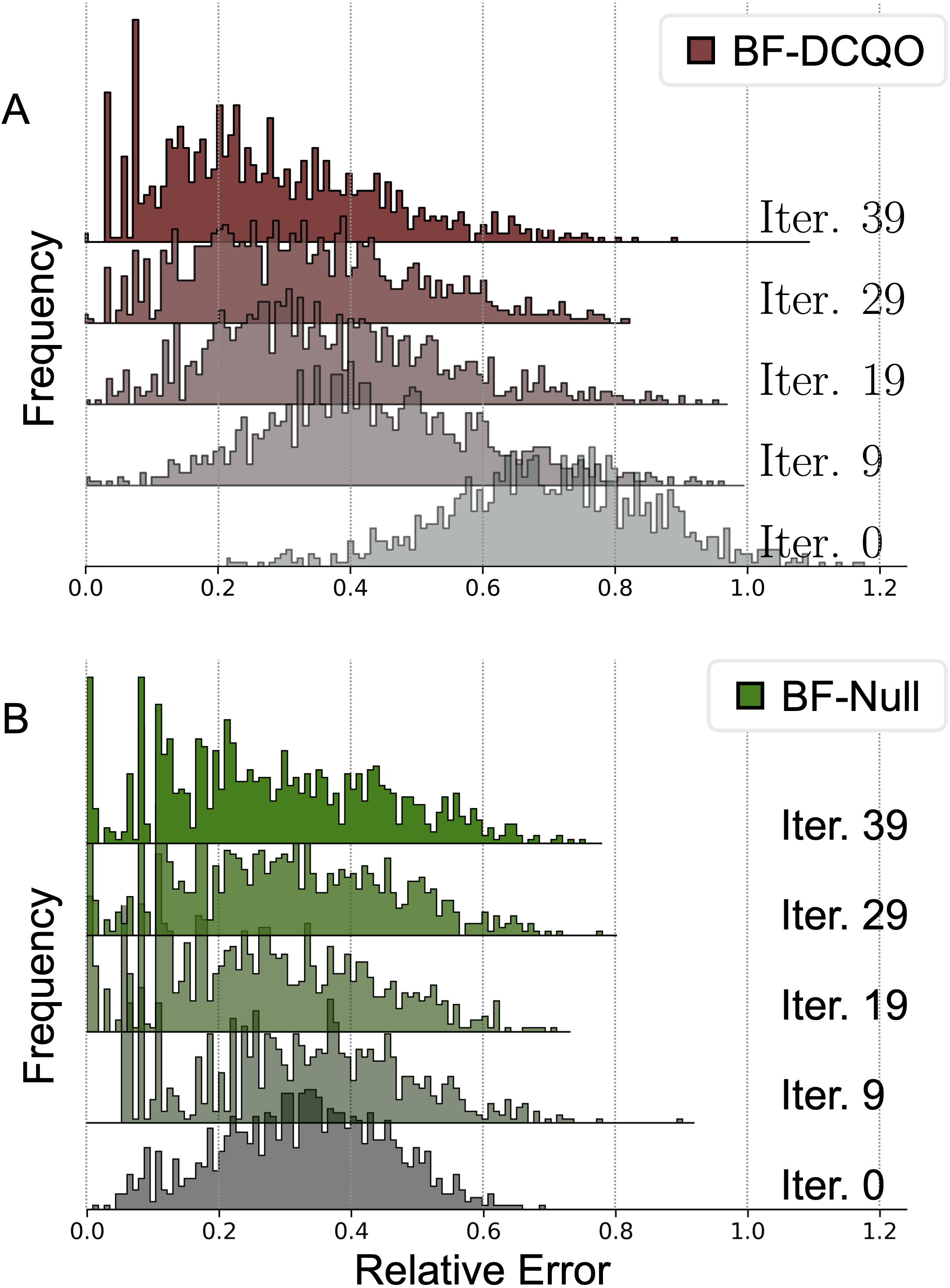} 
  \caption{Solution quality of 1000 reads for two random 29 qubit instances over multiple bias field iterations. (A) Distributions from a $t=2$ BF-DCQO run of the IonQ Forte noisy simulator taken from FIG. 1D in \cite{Cadavid_2025}. (B) Distributions from BF-Null.} 
  \label{fig:distB} 
\end{figure}

\section{Results}
\label{sec:Results}

\subsection{Ising problems}
\label{sec:Ising}

Following the approach in \cite{Cadavid_2025}, we benchmarked ensembles of 400 random instances of the fully connected Ising model $H(\vec{s}) = \sum_{i=1}^{N}h_js_i + \sum_{i<j}^{N}J_{ij}s_is_j$ where $s_i$ can
take values 1 and -1. The linear and quadratic weights $h_i$ and $J_{ij}$ were drawn randomly from a Gaussian distribution with mean zero and variance one.

The largest of these problems contains 29 variables. Since D-Wave quantum computers contain over 4400 qubits, we were able to embed each problem multiple times into the Advantage2\texttrademark~processor with the Advantage2\_system1 solver used in all of our tests  (see Section \ref{subsec:minor}). We performed a single programming per instance, collecting 1000 reads of $\SI{500}{\micro s}$ annealing time each (see Section~\ref{sec:timing}). 
In Figure~\ref{fig:p_gs}, we show the ground-state probability when embedding a single copy of the instance into the QPU (D-Wave Single),
and when embedding multiple copies in parallel and reporting the best solution per read (D-Wave Parallel), see Section~\ref{subsec:minor}.

The BF-DCQO and DCQO results in Figure~\ref{fig:p_gs} are reported in FIG. 1D in Ref.~\cite{Cadavid_2025}. BF-DCQO ran for 10 iterations on a noiseless simulator using $t=3$ Trotter steps and 1000 reads. DCQO is equivalent to a single iteration of BF-DCQO without the bias-fields \cite{Hegade_2022}. By analogy to BF-DCQO, the BF-Null results come from 10 bias-field iterations of 1000 reads, with constant $\alpha=0.02$  \cite{Romero2025, simen, chandarana,romero} and $\gamma=3$ (see Section \ref{subsec:bfnull}). 

Figure~\ref{fig:p_gs} shows that D-Wave Single and Parallel return ground-state probabilities that are significantly higher than BF-DCQO and DCQO, and the performance gap increases with problem size. We find that BF-Null shows a similar scaling to its hybrid counterpart BF-DCQO, but with a consistently higher ground-state probability. These experiments suggest that the performance of BF-DCQO is inferior to its trivial classical counterpart BF-Null.

Figure~\ref{fig:distA} shows the solution quality, measured as relative error (see Section \ref{subsec:relerr}), of 1000 reads from each solver. Figure~\ref{fig:distA}A displays the iterative evolution of a $t=2$ BF-DCQO run the IonQ Forte noisy simulator, for a single 29 qubit instance. This requires up to 39 sequential iterations of 1000 reads each. These results are reported in Figure 1D in \cite{Cadavid_2025}.

Figure~\ref{fig:distA}B shows the performance of the D-Wave QPU and BF-Null (iteration 39) on five random instances of the same problem class. The D-Wave results are from a single programming, 1000 reads and $\SI{500}{\micro s}$ annealing time. D-Wave solutions are significantly better than both BF solvers without using any iteration procedures, routinely returning optimal and near-optimal results. The BF-Null distribution appears shrunken due to sharing scale with the sharper distributions from the D-Wave QPU; we therefore highlight the mean and maximum of the distribution with vertical lines. 

Figure~\ref{fig:distB} provides a more detailed comparison of the BF solver distributions. We find that BF-Null performs significantly better than BF-DCQO at the first iteration, when the bias fields are empty. This suggests that
the performance of the trivial classical subsolver is superior to the BF-DCQO's DCQO subsolver, even when run on a simulator. Nevertheless, regardless of the difference in the subsolver performance, both methods seem to plateau at similar solution-quality levels instead of converging towards a high probability of optimal solutions. This suggests that the iterative BF mechanism may trap heuristics at similar-quality local minima.

\subsection{Hising problems}
\label{hising}

In this section we focus on higher-order Ising  (hising) problems, also known as higher-order unconstrained binary optimization (HUBO) problems: in particular, the cubic hising Hamiltonian $H(\vec{s}) = \sum_{i=1}^{N}h_is_i + \sum_{i<j}^{N}J_{ij}s_is_j + \sum_{i<j<k}^{N}K_{ijk}s_is_js_k$. Introducing cubic interactions into the Hamiltonian requires additional gates when implemented with a gate-model quantum computer and additional qubits when implemented with an annealing quantum computer.

\begin{figure}[t!] 
  \centering 
  \includegraphics[width=\columnwidth]{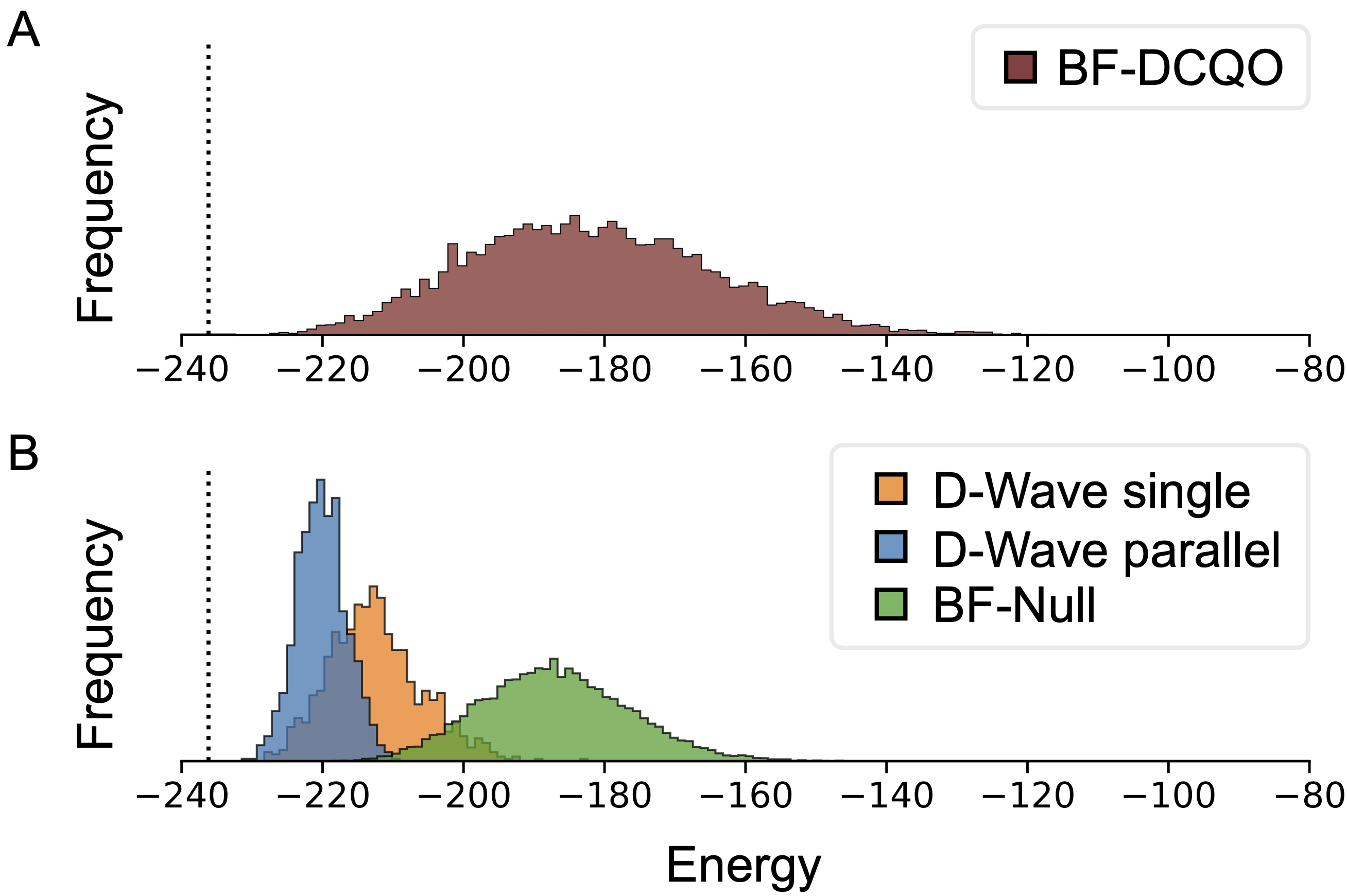} 
  \caption{Solution quality of the nearest-neighbour heavy hex hising problem with Gaussian weights. The vertical dashed line marks the problem's optimal solution (-236.18). (A) Tenth iteration of BF-DCQO running on ibm\_fez reported in Figure 2 in \cite{Romero2025}. (B) Solution qualities for a D-Wave quantum annealer using single and parallel embeddings and BF-Null (10 iterations).} 
  \label{fig:hising} 
\end{figure}

Contrary to its description in the paper \cite{Romero2025}, this hising problem does not ``perfectly match the IBM heavy-hex topology [...] avoiding an overhead of entangling gates for coupling distant qubits''. Instead, the instance is defined on a one-dimensional chain of 156 qubits, which is not a subgraph of the a 156 qubit heavy-hex lattice. In our recent communications with the authors, they clarified that long-range entangling SWAP gates were used, although many ended up discarded by a gate-cutoff threshold, potentially isolating qubits in the chain.

In our study, when mapping the problem into a D-Wave quantum annealer, we first reduce the problem to Ising (see Section~\ref{hising_to_ising}) and then we leverage the larger qubit count to minor-embed seven copies of the problem into one hardware graph and solve them in parallel during each anneal  (see Section~\ref{subsec:minor}). 

Figure~\ref{fig:hising} shows the solution quality of the solvers in these types of instances. The results in Figure~\ref{fig:hising}A are reported in Figure 2 in Ref.~\cite{Romero2025}. These $t=3$ BF-DCQO results correspond to the tenth iteration of the algorithm run on ibm\_fez. The previous iterations were simulated using IBM's MPS simulator under noise-free conditions. BF-DCQO uses 10000 reads per iteration, with $\alpha=0.02$. 

In Figure~\ref{fig:hising}B, the D-Wave quantum annealer results come from a single programming per problem, 1000 reads and $\SI{350}{\micro s}$ anneals (see Section~\ref{sec:timing}). We see that the QA results are notably better than the results from BF-DCQO using far less resources (see Section \ref{sec:timing}). Similarly to BF-DCQO, the BF-Null results come from the tenth iteration, with $\alpha=0.02$, $\gamma=2$, and 10000 reads. BF-Null distributions show a remarkable similarity with that of its quantum-hybrid counterpart BF-DQCO.

We note that Romero et al. \cite{Romero2025} report a distribution of QA solutions far worse than any distribution in Figure~\ref{fig:hising}. The authors mention using an older D-Wave QPU generation with default parameters. Nevertheless, in our own tests with the same QPU and parameters, we still find that the D-Wave quantum annealer returns better-quality solutions than those reported in \cite{Romero2025}. This is consistent with previous studies that report that D-Wave quantum annealers consistently return optimal and near-optimal results for heavy-hex hising models \cite{elijah127,elijahqaoa}.

\section{Timing}
\label{sec:timing}

In a recent paper \cite{mcgeoch2024}, McGeoch lays out simple rules for avoiding common pitfalls when benchmarking quantum computers. The first one is: \textit{Don't claim superior performance without mentioning runtimes}. Algorithmic performance research asks the
foundational question of how much time is needed to obtain a target solution quality. If time is not taken into account, brute force methods would win.

The time required to run a quantum annealer (called QPU access time) includes a one-time programming overhead, the annealing times, readouts, and delay between reads. In addition, if the problem graph is not a subgraph of a QPU, one needs to find a graph minor embedding (see Section~\ref{subsec:minor}), this constitutes a one-time cost per problem ensemble and is negligible as a fraction of total experimentation time.

In our tests of Ising instances described in \cite{Cadavid_2025}, shown in Figure~\ref{fig:p_gs}, we performed a single programming (35~ms) and 1000 anneals of $\SI{500}{\micro s}$  each, with $\SI{98}{\micro s}$  readout and $\SI{60}{\micro s}$  delay, totaling 693~ms of QPU access time on the Advantage2\_system1 solver. While longer annealing times tend to reach lower solution energies and higher success probabilities, we chose $\SI{500}{\micro s}$ anneals rather than the solver's $\SI{2000}{\micro s}$ maximum in order to fit 1000 reads within the solver's QPU access time limit of one second.

We do not have access to the timings of the simulators used in Ref.~\cite{Cadavid_2025}. However,  if BF-DQCO were to be implemented on an IonQ Forte platform, timing results from Wang et al. \cite{Wang_2024} suggest that execution times per gate are $\SI{970}{\micro s}$. Assuming 10 iterations and 1000 shots, a single-depth circuit would require 9.7~s, which is about 14x higher than D-Wave QPU access time for the full computation. This calculation ignores programming and classical processing overheads that would be required in any iterative scheme. We don't know the depth of the BF-DCQO circuit in IonQ Forte, but based on the Supplementary Materials \cite{Cadavid_2025} we estimate it to be at least 10, resulting in runtimes over 140$\times$ longer than D-Wave QPU access time.

For the hising instances described in Ref.~\cite{Romero2025}, we compare Advantage2\_system1 to BF-DCQO running on ibm\_fez for a tenth iteration after running the previous iterations on a noiseless MPS simulator.  

The D-Wave quantum annealing results come from a single programming (35~ms) and 1000 anneals of $\SI{350}{\micro s}$  each, with $\SI{98}{\micro s}$  readout and $\SI{60}{\micro s}$  delay, totaling 543~ms of QPU access time. We chose the $\SI{350}{\micro s}$  annealing time because it is the same time-per-shot that IBM reports, providing a similar cost per read. However, while all the reads from the D-Wave quantum annealer count as final, the BF-DCQO method goes through a number of iterations, and only the last set of samples count as final. 

IBM documentation \cite{workload} suggests the following rule for calculating quantum computation time in seconds: \textit{If you aren't using any advanced error-mitigation techniques or custom rep\_delay, you can use $2+0.00035\times<$num\_executions$>$ as a quick formula}. Consequently, a single iteration of BF-DCQO with 1000 reads would take $\sim2350$~ms and the whole BF-DQCO procedure with 10 iterations would take $\sim23500$~ms. That is about 43$\times$ longer than QA ignoring any processing overheads of the iterative scheme. We note these estimates are based on BF-DCQO operating with 1000 reads instead of the 10000 reads from the experiment in \cite{Cadavid_2025}. The experiment from the paper takes an estimated time of $\sim55$ seconds, about 101$\times$ longer than D-Wave QPU access time.

Finally, the following table compares QA and BF-DCQO times to the sampling times of BF-Null running on an Intel Core i9-7900X CPU for Ising problems and an AMD EPYC 9534 CPU with an NVIDIA L4 GPU for hising problems.

\begin{table}[h!]
  \begin{center}
    \begin{tabular}{ |c|c|c|c| }
      \hline
      \multicolumn{4}{|c|}{Runtime (seconds) for 1000 reads} \\
      \hline
      Experiment   & BF-Null & D-Wave&  BF-DCQO (est.)\\ 
      \hline
      Ising, Figure~\ref{fig:p_gs}   & 0.008 & 0.696  & $>97$\\ 
      \hline
      Ising, Figures~\ref{fig:distA},\ref{fig:distB} & 0.067 & 0.696   & $>378$ \\ 
      \hline
      Hising, Figure~\ref{fig:hising} & 0.032 & 0.543   & $>23.5$\\ 
      \hline
    \end{tabular}
    \caption{Timings for BF-Null and quantum annealing, and estimated timings for BF-DCQO.}
  \end{center}
\end{table}

\section{Conclusion}

These results show that D-Wave quantum annealers significantly outperform BF-DCQO in a variety of optimization problems by routinely finding better results in orders-of-magnitude shorter time. This is consistent with current literature, where annealing quantum computers clearly outperform other quantum systems in optimization tasks \cite{rigettiwhitepaper,Willsch_2020, lubinski, elijah127,elijahqaoa, qaoawhitepaper}.

We also emphasize the importance of both reporting computation times when evaluating optimization performance as well as disclosing sufficient details in test designs to allow independent replication \cite{mcgeoch2024}.

\section{Methods}

\subsection{Bias-field null-hypothesis}
\label{subsec:bfnull}

The bias-field null-hypothesis (BF-Null) method consists in replacing the quantum computer in BF-DCQO with a simple classical routine as a subsolver. 
This type of sanity check to detect a quantum contribution can be found in previous performance studies of hybrid algorithms  (e.g. \cite{raymond}).

Here we use a simple single zero-temperature sweep subsolver: starting from a random
solution, pass once through all variables in fixed order and flip their state if total solution energy is improved. 
In general, this produces slightly better than random solutions as it only considers $N+1$ different states in a $2^N$ solution landscape.
The algorithm is deterministic except for the initial random solution, and the complexity is $O(N+M+K)$, where $N$, $M$, $K$ are the number of linear, quadratic and three-body terms in the graph, respectively. 

The complete procedure consists of performing $b$ bias-field iterations with the subsolver (Alg.\ref{alg:bfnh}). At each iteration, the objective
function includes additional linear terms proportional to the average of the $\alpha$-best solutions in the previous iteration \cite{Romero2025, simen, chandarana,romero}.

\renewcommand{\algorithmicrequire}{\textbf{Input:}}
\renewcommand{\algorithmicensure}{\textbf{Output:}}

\begin{algorithm}
  \caption{Bias-field null-hypothesis}\label{alg:bfnh}
  \begin{algorithmic}[h!]
  \Require Cost function $E(\vec{s})$, number of bias-field updates $b$, number of reads $R$, sample selection $0<\alpha<1$, bias-field weight $\gamma$
  \Ensure R states $\vec{s}$
  \State $\vec{B} \gets \vec{0}$ \Comment{Initialize bias fields}
  \For{$i=1$ to $b$}
    \State $E'(\vec{s}) = E(\vec{s}) + \gamma \vec{B} \cdot \vec{s}$
    \State $S \gets \emptyset$ \Comment{Initialize sample set} 
    \For{$j=1$ to $R$}
      \State Generate random solution  $\vec{s}$
      \For{$s_k$ in $\vec{s}$}
        \State $s_k = -s_k$ if $E'(s_k|\vec{s}) > E'(-s_k|\vec{s})$  
      \EndFor
      \State Add $\vec{s}$ to $S$
    \EndFor
  \State $\vec{B} \gets -<\vec{s}>_{\alpha}$ \Comment{Update $\vec{B}$ based on best samples} 
  \EndFor
  \State
  \Return S 

  \end{algorithmic}
\end{algorithm}

\subsection{Relative error}
\label{subsec:relerr}

These experiments are based on finding low energy states of a problem Hamiltonian. The optimal solution of a problem is an assignment of binary variables that has an energy called the ground-state energy  $e_{\rm gs}$. In this work, we find exact ground states with the \textit{TreeDecompositionSolver} available in the open-source D-Wave Ocean\texttrademark~SDK \cite{ocean}.

Since different random instances have different $e_{\rm gs}$, we compare solution qualities across instances based on relative error: the normalized distance between optimal $e$ and $e_{\rm gs}$,

\begin{equation}
  RE = \frac{e-e_{\rm gs}}{|e_{\rm gs}|},
\end{equation}
where $RE=0$ corresponds to optimality.

In Figures \ref{fig:distA}A and \ref{fig:distB}A, adapted from Ref.~\cite{Cadavid_2025}, the lowest energy on the left corresponds to the ground state $e_{\rm gs}$, with $RE=0$. The $e=0$ position in the x-axis corresponds to $RE=1$ when $e_{\rm gs}$ is negative. 

We note that the reported ground-state energy of the Ising instance shown in Figures \ref{fig:distA}A, \ref{fig:distB}A is statistically consistent with the ground states of our generated ensemble of instances used in Figures \ref{fig:distA}B, \ref{fig:distB}B.

\subsection{Minor embedding}
\label{subsec:minor}

Problems that are not subgraphs of a QPU's qubit layout use the process of minor-embedding, coupling multiple qubits in chains that represent single variables. One may embed in parallel multiple copies of one problem into the processor if they fit.

All embeddings in this paper were found using the \textit{minorminer.find\_embedding} tool based on \cite{cai2014practicalheuristicfindinggraph}, available in the open-source D-Wave Ocean SDK \cite{ocean}. The parallel embeddings in Table~\ref{tab:emb} were found by an iterative method that consists of embedding one copy of the problem at a time onto the unused qubits of the QPU until \textit{minorminer.find\_embedding} fails to fit more embeddings.

\begin{table}[h!]

  \begin{center}
    \begin{tabular}{ |c|c| }
      \hline
      \multicolumn{2}{|c|}{Number of parallel embeddings} \\
      \hline
      Problem class & Count \\ 
      \hline
      Clique N=10 in Figure~\ref{fig:p_gs}& 196 \\ 
      \hline
      Clique N=12 in Figure~\ref{fig:p_gs}& 150 \\ 
      \hline
      Clique N=14 in Figure~\ref{fig:p_gs}& 114 \\ 
      \hline
      Clique N=16 in Figure~\ref{fig:p_gs}& 93 \\ 
      \hline
      Clique N=18 in Figure~\ref{fig:p_gs}& 72 \\ 
      \hline
      Clique N=20 in Figure~\ref{fig:p_gs}& 57 \\ 
      \hline
      Clique N=29 in Figures~\ref{fig:distA},\ref{fig:distB}& 27 \\ 
      \hline
      ibm\_fez N=156 in Figure~\ref{fig:hising}& 7 \\ 
      \hline

    \end{tabular}
  \end{center}

\caption{Number of parallel embeddings fit into the Advantage2\_system1. \label{tab:emb}}

\end{table}

\subsection{Hising to Ising reduction}
\label{hising_to_ising}

The reduction of a hising problem to Ising requires representing products of variables with auxiliary variables, and then setting penalties to ensure that the value of the product variable is consistent with the values of the original variables.

In the experiments of Section~\ref{hising}, we reduced the three-body hising terms to pairwise Ising terms by introducing up to two auxiliary variables per cubic term, with a penalty equal to 5, using the \textit{dimod.higherorder.make\_quadratic} tool available in the D-Wave Ocean SDK \cite{ocean}.

\subsection{Data availability}

Supporting problem instances, solver solutions and ground-state energies are publicly available at the Zenodo repository \url{https://doi.org/10.5281/zenodo.23376568}

\section{Acknowledgements}

We thank Andrew~D.~King, Trevor~Lanting, Joel~Pasvolsky, Jack~Raymond and Richard Harris for fruitful discussions and comments on the manuscript.

\bibliography{paper}

\begin{thebibliography}{21}%
\makeatletter
\providecommand \@ifxundefined [1]{%
 \@ifx{#1\undefined}
}%
\providecommand \@ifnum [1]{%
 \ifnum #1\expandafter \@firstoftwo
 \else \expandafter \@secondoftwo
 \fi
}%
\providecommand \@ifx [1]{%
 \ifx #1\expandafter \@firstoftwo
 \else \expandafter \@secondoftwo
 \fi
}%
\providecommand \natexlab [1]{#1}%
\providecommand \enquote  [1]{``#1''}%
\providecommand \bibnamefont  [1]{#1}%
\providecommand \bibfnamefont [1]{#1}%
\providecommand \citenamefont [1]{#1}%
\providecommand \href@noop [0]{\@secondoftwo}%
\providecommand \href [0]{\begingroup \@sanitize@url \@href}%
\providecommand \@href[1]{\@@startlink{#1}\@@href}%
\providecommand \@@href[1]{\endgroup#1\@@endlink}%
\providecommand \@sanitize@url [0]{\catcode `\\12\catcode `\$12\catcode
  `\&12\catcode `\#12\catcode `\^12\catcode `\_12\catcode `\%12\relax}%
\providecommand \@@startlink[1]{}%
\providecommand \@@endlink[0]{}%
\providecommand \url  [0]{\begingroup\@sanitize@url \@url }%
\providecommand \@url [1]{\endgroup\@href {#1}{\urlprefix }}%
\providecommand \urlprefix  [0]{URL }%
\providecommand \Eprint [0]{\href }%
\providecommand \doibase [0]{https://doi.org/}%
\providecommand \selectlanguage [0]{\@gobble}%
\providecommand \bibinfo  [0]{\@secondoftwo}%
\providecommand \bibfield  [0]{\@secondoftwo}%
\providecommand \translation [1]{[#1]}%
\providecommand \BibitemOpen [0]{}%
\providecommand \bibitemStop [0]{}%
\providecommand \bibitemNoStop [0]{.\EOS\space}%
\providecommand \EOS [0]{\spacefactor3000\relax}%
\providecommand \BibitemShut  [1]{\csname bibitem#1\endcsname}%
\let\auto@bib@innerbib\@empty
\bibitem [{\citenamefont {Romero}\ \emph
  {et~al.}(2025{\natexlab{a}})\citenamefont {Romero}, \citenamefont {Visuri},
  \citenamefont {Cadavid}, \citenamefont {Simen}, \citenamefont {Solano},\ and\
  \citenamefont {Hegade}}]{Romero2025}%
  \BibitemOpen
  \bibfield  {author} {\bibinfo {author} {\bibfnamefont {S.~V.}\ \bibnamefont
  {Romero}}, \bibinfo {author} {\bibfnamefont {A.-M.}\ \bibnamefont {Visuri}},
  \bibinfo {author} {\bibfnamefont {A.~G.}\ \bibnamefont {Cadavid}}, \bibinfo
  {author} {\bibfnamefont {A.}~\bibnamefont {Simen}}, \bibinfo {author}
  {\bibfnamefont {E.}~\bibnamefont {Solano}},\ and\ \bibinfo {author}
  {\bibfnamefont {N.~N.}\ \bibnamefont {Hegade}},\ }\bibfield  {title}
  {\bibinfo {title} {Bias-field digitized counterdiabatic quantum algorithm for
  higher-order binary optimization},\ }\href
  {https://doi.org/10.1038/s42005-025-02270-3} {\bibfield  {journal} {\bibinfo
  {journal} {Communications Physics}\ }\textbf {\bibinfo {volume} {8}},\
  \bibinfo {pages} {348} (\bibinfo {year} {2025}{\natexlab{a}})}\BibitemShut
  {NoStop}%
\bibitem [{\citenamefont {Cadavid}\ \emph {et~al.}(2025)\citenamefont
  {Cadavid}, \citenamefont {Dalal}, \citenamefont {Simen}, \citenamefont
  {Solano},\ and\ \citenamefont {Hegade}}]{Cadavid_2025}%
  \BibitemOpen
  \bibfield  {author} {\bibinfo {author} {\bibfnamefont {A.~G.}\ \bibnamefont
  {Cadavid}}, \bibinfo {author} {\bibfnamefont {A.}~\bibnamefont {Dalal}},
  \bibinfo {author} {\bibfnamefont {A.}~\bibnamefont {Simen}}, \bibinfo
  {author} {\bibfnamefont {E.}~\bibnamefont {Solano}},\ and\ \bibinfo {author}
  {\bibfnamefont {N.~N.}\ \bibnamefont {Hegade}},\ }\bibfield  {title}
  {\bibinfo {title} {Bias-field digitized counterdiabatic quantum
  optimization},\ }\bibfield  {journal} {\bibinfo  {journal} {Physical Review
  Research}\ }\textbf {\bibinfo {volume} {7}},\ \href
  {https://doi.org/10.1103/physrevresearch.7.l022010}
  {10.1103/physrevresearch.7.l022010} (\bibinfo {year} {2025})\BibitemShut
  {NoStop}%
\bibitem [{\citenamefont {Kadowaki}\ and\ \citenamefont
  {Nishimori}(1998)}]{kadowaki}%
  \BibitemOpen
  \bibfield  {author} {\bibinfo {author} {\bibfnamefont {T.}~\bibnamefont
  {Kadowaki}}\ and\ \bibinfo {author} {\bibfnamefont {H.}~\bibnamefont
  {Nishimori}},\ }\bibfield  {title} {\bibinfo {title} {Quantum annealing in
  the transverse ising model},\ }\href
  {https://doi.org/10.1103/PhysRevE.58.5355} {\bibfield  {journal} {\bibinfo
  {journal} {Phys. Rev. E}\ }\textbf {\bibinfo {volume} {58}},\ \bibinfo
  {pages} {5355} (\bibinfo {year} {1998})}\BibitemShut {NoStop}%
\bibitem [{\citenamefont {Farhi}\ \emph {et~al.}(2000)\citenamefont {Farhi},
  \citenamefont {Goldstone}, \citenamefont {Gutmann},\ and\ \citenamefont
  {Sipser}}]{farhi2000}%
  \BibitemOpen
  \bibfield  {author} {\bibinfo {author} {\bibfnamefont {E.}~\bibnamefont
  {Farhi}}, \bibinfo {author} {\bibfnamefont {J.}~\bibnamefont {Goldstone}},
  \bibinfo {author} {\bibfnamefont {S.}~\bibnamefont {Gutmann}},\ and\ \bibinfo
  {author} {\bibfnamefont {M.}~\bibnamefont {Sipser}},\ }\href
  {https://arxiv.org/abs/quant-ph/0001106} {\bibinfo {title} {Quantum
  computation by adiabatic evolution}} (\bibinfo {year} {2000}),\ \Eprint
  {https://arxiv.org/abs/quant-ph/0001106} {arXiv:quant-ph/0001106 [quant-ph]}
  \BibitemShut {NoStop}%
\bibitem [{\citenamefont {Farhi}\ \emph {et~al.}(2014)\citenamefont {Farhi},
  \citenamefont {Goldstone},\ and\ \citenamefont {Gutmann}}]{farhi2014}%
  \BibitemOpen
  \bibfield  {author} {\bibinfo {author} {\bibfnamefont {E.}~\bibnamefont
  {Farhi}}, \bibinfo {author} {\bibfnamefont {J.}~\bibnamefont {Goldstone}},\
  and\ \bibinfo {author} {\bibfnamefont {S.}~\bibnamefont {Gutmann}},\ }\href
  {https://arxiv.org/abs/1411.4028} {\bibinfo {title} {A quantum approximate
  optimization algorithm}} (\bibinfo {year} {2014}),\ \Eprint
  {https://arxiv.org/abs/1411.4028} {arXiv:1411.4028 [quant-ph]} \BibitemShut
  {NoStop}%
\bibitem [{\citenamefont {Hegade}\ \emph {et~al.}(2022)\citenamefont {Hegade},
  \citenamefont {Chen},\ and\ \citenamefont {Solano}}]{Hegade_2022}%
  \BibitemOpen
  \bibfield  {author} {\bibinfo {author} {\bibfnamefont {N.~N.}\ \bibnamefont
  {Hegade}}, \bibinfo {author} {\bibfnamefont {X.}~\bibnamefont {Chen}},\ and\
  \bibinfo {author} {\bibfnamefont {E.}~\bibnamefont {Solano}},\ }\bibfield
  {title} {\bibinfo {title} {Digitized counterdiabatic quantum optimization},\
  }\bibfield  {journal} {\bibinfo  {journal} {Physical Review Research}\
  }\textbf {\bibinfo {volume} {4}},\ \href
  {https://doi.org/10.1103/physrevresearch.4.l042030}
  {10.1103/physrevresearch.4.l042030} (\bibinfo {year} {2022})\BibitemShut
  {NoStop}%
\bibitem [{\citenamefont {Simen}\ \emph {et~al.}(2025)\citenamefont {Simen},
  \citenamefont {Romero}, \citenamefont {Cadavid}, \citenamefont {Solano},\
  and\ \citenamefont {Hegade}}]{simen}%
  \BibitemOpen
  \bibfield  {author} {\bibinfo {author} {\bibfnamefont {A.}~\bibnamefont
  {Simen}}, \bibinfo {author} {\bibfnamefont {S.~V.}\ \bibnamefont {Romero}},
  \bibinfo {author} {\bibfnamefont {A.~G.}\ \bibnamefont {Cadavid}}, \bibinfo
  {author} {\bibfnamefont {E.}~\bibnamefont {Solano}},\ and\ \bibinfo {author}
  {\bibfnamefont {N.~N.}\ \bibnamefont {Hegade}},\ }\href
  {https://arxiv.org/abs/2504.15367} {\bibinfo {title} {Branch-and-bound
  digitized counterdiabatic quantum optimization}} (\bibinfo {year} {2025}),\
  \Eprint {https://arxiv.org/abs/2504.15367} {arXiv:2504.15367 [quant-ph]}
  \BibitemShut {NoStop}%
\bibitem [{\citenamefont {Chandarana}\ \emph {et~al.}(2025)\citenamefont
  {Chandarana}, \citenamefont {Cadavid}, \citenamefont {Romero}, \citenamefont
  {Simen}, \citenamefont {Solano},\ and\ \citenamefont {Hegade}}]{chandarana}%
  \BibitemOpen
  \bibfield  {author} {\bibinfo {author} {\bibfnamefont {P.}~\bibnamefont
  {Chandarana}}, \bibinfo {author} {\bibfnamefont {A.~G.}\ \bibnamefont
  {Cadavid}}, \bibinfo {author} {\bibfnamefont {S.~V.}\ \bibnamefont {Romero}},
  \bibinfo {author} {\bibfnamefont {A.}~\bibnamefont {Simen}}, \bibinfo
  {author} {\bibfnamefont {E.}~\bibnamefont {Solano}},\ and\ \bibinfo {author}
  {\bibfnamefont {N.~N.}\ \bibnamefont {Hegade}},\ }\href
  {https://arxiv.org/abs/2505.08663} {\bibinfo {title} {Runtime quantum
  advantage with digital quantum optimization}} (\bibinfo {year} {2025}),\
  \Eprint {https://arxiv.org/abs/2505.08663} {arXiv:2505.08663 [quant-ph]}
  \BibitemShut {NoStop}%
\bibitem [{\citenamefont {Romero}\ \emph
  {et~al.}(2025{\natexlab{b}})\citenamefont {Romero}, \citenamefont {Cadavid},
  \citenamefont {Nikačević}, \citenamefont {Solano}, \citenamefont {Hegade},
  \citenamefont {Lopez-Ruiz}, \citenamefont {Girotto}, \citenamefont {Yamada},
  \citenamefont {Barkoutsos}, \citenamefont {Kaushik},\ and\ \citenamefont
  {Roetteler}}]{romero}%
  \BibitemOpen
  \bibfield  {author} {\bibinfo {author} {\bibfnamefont {S.~V.}\ \bibnamefont
  {Romero}}, \bibinfo {author} {\bibfnamefont {A.~G.}\ \bibnamefont {Cadavid}},
  \bibinfo {author} {\bibfnamefont {P.}~\bibnamefont {Nikačević}}, \bibinfo
  {author} {\bibfnamefont {E.}~\bibnamefont {Solano}}, \bibinfo {author}
  {\bibfnamefont {N.~N.}\ \bibnamefont {Hegade}}, \bibinfo {author}
  {\bibfnamefont {M.~A.}\ \bibnamefont {Lopez-Ruiz}}, \bibinfo {author}
  {\bibfnamefont {C.}~\bibnamefont {Girotto}}, \bibinfo {author} {\bibfnamefont
  {M.}~\bibnamefont {Yamada}}, \bibinfo {author} {\bibfnamefont {P.~K.}\
  \bibnamefont {Barkoutsos}}, \bibinfo {author} {\bibfnamefont
  {A.}~\bibnamefont {Kaushik}},\ and\ \bibinfo {author} {\bibfnamefont
  {M.}~\bibnamefont {Roetteler}},\ }\href {https://arxiv.org/abs/2506.07866}
  {\bibinfo {title} {Protein folding with an all-to-all trapped-ion quantum
  computer}} (\bibinfo {year} {2025}{\natexlab{b}}),\ \Eprint
  {https://arxiv.org/abs/2506.07866} {arXiv:2506.07866 [quant-ph]} \BibitemShut
  {NoStop}%
\bibitem [{\citenamefont {Pelofske}\ \emph {et~al.}(2023)\citenamefont
  {Pelofske}, \citenamefont {B\"{a}rtschi},\ and\ \citenamefont
  {Eidenbenz}}]{elijah127}%
  \BibitemOpen
  \bibfield  {author} {\bibinfo {author} {\bibfnamefont {E.}~\bibnamefont
  {Pelofske}}, \bibinfo {author} {\bibfnamefont {A.}~\bibnamefont
  {B\"{a}rtschi}},\ and\ \bibinfo {author} {\bibfnamefont {S.}~\bibnamefont
  {Eidenbenz}},\ }\bibfield  {title} {\bibinfo {title} {Quantum annealing vs.
  qaoa: 127 qubit higher-order ising problems on nisq computers},\ }in\ \href
  {https://doi.org/10.1007/978-3-031-32041-5_13} {\emph {\bibinfo {booktitle}
  {High Performance Computing: 38th International Conference, ISC High
  Performance 2023, Hamburg, Germany, May 21–25, 2023, Proceedings}}}\
  (\bibinfo  {publisher} {Springer-Verlag},\ \bibinfo {address} {Berlin,
  Heidelberg},\ \bibinfo {year} {2023})\ p.\ \bibinfo {pages}
  {240–258}\BibitemShut {NoStop}%
\bibitem [{\citenamefont {Pelofske}\ \emph {et~al.}(2024)\citenamefont
  {Pelofske}, \citenamefont {B{\"a}rtschi},\ and\ \citenamefont
  {Eidenbenz}}]{elijahqaoa}%
  \BibitemOpen
  \bibfield  {author} {\bibinfo {author} {\bibfnamefont {E.}~\bibnamefont
  {Pelofske}}, \bibinfo {author} {\bibfnamefont {A.}~\bibnamefont
  {B{\"a}rtschi}},\ and\ \bibinfo {author} {\bibfnamefont {S.}~\bibnamefont
  {Eidenbenz}},\ }\bibfield  {title} {\bibinfo {title} {Short-depth qaoa
  circuits and quantum annealing on higher-order ising models},\ }\href
  {https://doi.org/10.1038/s41534-024-00825-w} {\bibfield  {journal} {\bibinfo
  {journal} {npj Quantum Information}\ }\textbf {\bibinfo {volume} {10}},\
  \bibinfo {pages} {30} (\bibinfo {year} {2024})}\BibitemShut {NoStop}%
\bibitem [{\citenamefont {McGeoch}(2024)}]{mcgeoch2024}%
  \BibitemOpen
  \bibfield  {author} {\bibinfo {author} {\bibfnamefont {C.}~\bibnamefont
  {McGeoch}},\ }\href {https://arxiv.org/abs/2411.08860} {\bibinfo {title} {How
  not to fool the masses when giving performance results for quantum
  computers}} (\bibinfo {year} {2024}),\ \Eprint
  {https://arxiv.org/abs/2411.08860} {arXiv:2411.08860 [quant-ph]} \BibitemShut
  {NoStop}%
\bibitem [{\citenamefont {Wang}\ \emph {et~al.}(2024)\citenamefont {Wang},
  \citenamefont {Das},\ and\ \citenamefont {Nair}}]{Wang_2024}%
  \BibitemOpen
  \bibfield  {author} {\bibinfo {author} {\bibfnamefont {M.}~\bibnamefont
  {Wang}}, \bibinfo {author} {\bibfnamefont {P.}~\bibnamefont {Das}},\ and\
  \bibinfo {author} {\bibfnamefont {P.~J.}\ \bibnamefont {Nair}},\ }\bibfield
  {title} {\bibinfo {title} {Qoncord: A multi-device job scheduling framework
  for variational quantum algorithms},\ }in\ \href
  {https://doi.org/10.1109/micro61859.2024.00060} {\emph {\bibinfo {booktitle}
  {2024 57th IEEE/ACM International Symposium on Microarchitecture (MICRO)}}}\
  (\bibinfo  {publisher} {IEEE},\ \bibinfo {year} {2024})\ p.\ \bibinfo {pages}
  {735–749}\BibitemShut {NoStop}%
\bibitem [{wor()}]{workload}%
  \BibitemOpen
  \href@noop {} {\bibinfo {title} {{IBM~Quantum~Platform}:~workload usage}},\
  \bibinfo {howpublished}
  {\url{https://quantum.cloud.ibm.com/docs/en/guides/estimate-job-run-time}},\
  \bibinfo {note} {accessed: 2025-08-13}\BibitemShut {NoStop}%
\bibitem [{rig(2018)}]{rigettiwhitepaper}%
  \BibitemOpen
  \href@noop {} {\emph {\bibinfo {title} {{A Head-to-Head Comparison of D-Wave
  and Rigetti QPUs}}}},\ \bibinfo {number} {4-1025A-D}\ (\bibinfo  {publisher}
  {D-Wave Whitepaper},\ \bibinfo {year} {2018})\BibitemShut {NoStop}%
\bibitem [{\citenamefont {Willsch}\ \emph {et~al.}(2020)\citenamefont
  {Willsch}, \citenamefont {Willsch}, \citenamefont {Jin}, \citenamefont
  {De~Raedt},\ and\ \citenamefont {Michielsen}}]{Willsch_2020}%
  \BibitemOpen
  \bibfield  {author} {\bibinfo {author} {\bibfnamefont {M.}~\bibnamefont
  {Willsch}}, \bibinfo {author} {\bibfnamefont {D.}~\bibnamefont {Willsch}},
  \bibinfo {author} {\bibfnamefont {F.}~\bibnamefont {Jin}}, \bibinfo {author}
  {\bibfnamefont {H.}~\bibnamefont {De~Raedt}},\ and\ \bibinfo {author}
  {\bibfnamefont {K.}~\bibnamefont {Michielsen}},\ }\bibfield  {title}
  {\bibinfo {title} {Benchmarking the quantum approximate optimization
  algorithm},\ }\bibfield  {journal} {\bibinfo  {journal} {Quantum Information
  Processing}\ }\textbf {\bibinfo {volume} {19}},\ \href
  {https://doi.org/10.1007/s11128-020-02692-8} {10.1007/s11128-020-02692-8}
  (\bibinfo {year} {2020})\BibitemShut {NoStop}%
\bibitem [{\citenamefont {Lubinski}\ \emph {et~al.}(2024)\citenamefont
  {Lubinski}, \citenamefont {Coffrin}, \citenamefont {McGeoch}, \citenamefont
  {Sathe}, \citenamefont {Apanavicius}, \citenamefont {Bernal~Neira},\ and\
  \citenamefont {Collaboration}}]{lubinski}%
  \BibitemOpen
  \bibfield  {author} {\bibinfo {author} {\bibfnamefont {T.}~\bibnamefont
  {Lubinski}}, \bibinfo {author} {\bibfnamefont {C.}~\bibnamefont {Coffrin}},
  \bibinfo {author} {\bibfnamefont {C.}~\bibnamefont {McGeoch}}, \bibinfo
  {author} {\bibfnamefont {P.}~\bibnamefont {Sathe}}, \bibinfo {author}
  {\bibfnamefont {J.}~\bibnamefont {Apanavicius}}, \bibinfo {author}
  {\bibfnamefont {D.}~\bibnamefont {Bernal~Neira}},\ and\ \bibinfo {author}
  {\bibfnamefont {Q.~E. D. C.-C.}\ \bibnamefont {Collaboration}},\ }\bibfield
  {title} {\bibinfo {title} {Optimization applications as quantum performance
  benchmarks},\ }\bibfield  {journal} {\bibinfo  {journal} {ACM Transactions on
  Quantum Computing}\ }\textbf {\bibinfo {volume} {5}},\ \href
  {https://doi.org/10.1145/3678184} {10.1145/3678184} (\bibinfo {year}
  {2024})\BibitemShut {NoStop}%
\bibitem [{qao(2023)}]{qaoawhitepaper}%
  \BibitemOpen
  \href@noop {} {\emph {\bibinfo {title} {{Optimization Performance of QA and
  QAOA}}}},\ \bibinfo {number} {14-0240-A}\ (\bibinfo  {publisher} {D-Wave
  Whitepaper},\ \bibinfo {year} {2023})\BibitemShut {NoStop}%
\bibitem [{\citenamefont {Raymond}\ \emph {et~al.}(2023)\citenamefont
  {Raymond}, \citenamefont {Stevanovic}, \citenamefont {Bernoudy},
  \citenamefont {Boothby}, \citenamefont {McGeoch}, \citenamefont {Berkley},
  \citenamefont {Farr\'{e}}, \citenamefont {Pasvolsky},\ and\ \citenamefont
  {King}}]{raymond}%
  \BibitemOpen
  \bibfield  {author} {\bibinfo {author} {\bibfnamefont {J.}~\bibnamefont
  {Raymond}}, \bibinfo {author} {\bibfnamefont {R.}~\bibnamefont {Stevanovic}},
  \bibinfo {author} {\bibfnamefont {W.}~\bibnamefont {Bernoudy}}, \bibinfo
  {author} {\bibfnamefont {K.}~\bibnamefont {Boothby}}, \bibinfo {author}
  {\bibfnamefont {C.~C.}\ \bibnamefont {McGeoch}}, \bibinfo {author}
  {\bibfnamefont {A.~J.}\ \bibnamefont {Berkley}}, \bibinfo {author}
  {\bibfnamefont {P.}~\bibnamefont {Farr\'{e}}}, \bibinfo {author}
  {\bibfnamefont {J.}~\bibnamefont {Pasvolsky}},\ and\ \bibinfo {author}
  {\bibfnamefont {A.~D.}\ \bibnamefont {King}},\ }\bibfield  {title} {\bibinfo
  {title} {Hybrid quantum annealing for larger-than-qpu lattice-structured
  problems},\ }\bibfield  {journal} {\bibinfo  {journal} {ACM Transactions on
  Quantum Computing}\ }\textbf {\bibinfo {volume} {4}},\ \href
  {https://doi.org/10.1145/3579368} {10.1145/3579368} (\bibinfo {year}
  {2023})\BibitemShut {NoStop}%
\bibitem [{oce()}]{ocean}%
  \BibitemOpen
  \href@noop {} {\bibinfo {title} {{D-Wave Ocean SDK}}},\ \bibinfo
  {howpublished} {\url{https://github.com/dwavesystems/dwave-ocean-sdk}},\
  \bibinfo {note} {accessed: 2025-08-27}\BibitemShut {NoStop}%
\bibitem [{\citenamefont {Cai}\ \emph {et~al.}(2014)\citenamefont {Cai},
  \citenamefont {Macready},\ and\ \citenamefont
  {Roy}}]{cai2014practicalheuristicfindinggraph}%
  \BibitemOpen
  \bibfield  {author} {\bibinfo {author} {\bibfnamefont {J.}~\bibnamefont
  {Cai}}, \bibinfo {author} {\bibfnamefont {W.~G.}\ \bibnamefont {Macready}},\
  and\ \bibinfo {author} {\bibfnamefont {A.}~\bibnamefont {Roy}},\ }\href
  {https://arxiv.org/abs/1406.2741} {\bibinfo {title} {A practical heuristic
  for finding graph minors}} (\bibinfo {year} {2014}),\ \Eprint
  {https://arxiv.org/abs/1406.2741} {arXiv:1406.2741 [quant-ph]} \BibitemShut
  {NoStop}%
\end{thebibliography}%

\end{document}